\documentclass[reprint,amsmath,amssymb,prb,groupedaddress,twocolumn,superscriptaddress, floatfix]{revtex4-2}

\usepackage{times}
\usepackage{changes}

\usepackage{graphicx}
\usepackage{amssymb}
\usepackage{epstopdf}
\usepackage{amsmath}
\usepackage{color}
\usepackage{bbold}
\usepackage{enumitem}
\usepackage[acronym]{glossaries}
\usepackage{physics}
\usepackage[bottom]{footmisc}

\usepackage{orcidlink}

\makeatletter

\@ifundefined{textcolor}{}
{
	\definecolor{BLACK}{gray}{0}
	\definecolor{WHITE}{gray}{1}
	\definecolor{RED}{rgb}{1,0,0}
	\definecolor{GREEN}{rgb}{0,1,0}
	\definecolor{BLUE}{rgb}{0,0,1}
	\definecolor{CYAN}{cmyk}{1,0,0,0}
	\definecolor{MAGENTA}{cmyk}{0,1,0,0}
	\definecolor{YELLOW}{cmyk}{0,0,1,0}
}

\graphicspath{{Images/}}

\hypersetup{
	pdfpagemode={UseOutlines},
	bookmarksopen,
	pdfstartview={FitH},
	colorlinks,
	linkcolor={black},
	citecolor={blue},
	urlcolor={blue}}

\usepackage{hyperref}
\hypersetup{
	colorlinks=true,
	linkcolor=black,          
	citecolor=blue,           
	filecolor=magenta         
}

\makeatother	

\begin{document}
	
	\title{Measurement-induced phase transition in a classical, chaotic many-body system}

	\author{Josef Willsher \orcidlink{0000-0002-6895-6039}}
	\email{joe.willsher@tum.de}
    \affiliation{Department of Physics TQM, Technische Universit{\"a}t M{\"u}nchen, James-Franck-Stra{\ss}e 1, 85748 Garching, Germany}
	\author{Shu-Wei Liu \orcidlink{0000-0003-2352-1776}}
	\affiliation{Max-Planck-Institut f{\"u}r Physik komplexer Systeme, N{\"o}thnitzer Stra{\ss}e 38, 01187 Dresden, Germany}
	\author{Roderich Moessner}
	\affiliation{Max-Planck-Institut f{\"u}r Physik komplexer Systeme, N{\"o}thnitzer Stra{\ss}e 38, 01187 Dresden, Germany}
	\author{Johannes Knolle \orcidlink{0000-0002-0956-2419}}
	\affiliation{Department of Physics TQM, Technische Universit{\"a}t M{\"u}nchen, James-Franck-Stra{\ss}e 1, 85748 Garching, Germany}
	\affiliation{Munich Center for Quantum Science and Technology (MCQST), 80799 Munich, Germany}
	\affiliation{Blackett Laboratory, Imperial College London, London SW7 2AZ, United Kingdom}
 
	\begin{abstract}
       Local measurements in quantum systems are projective operations which act to counteract the spread of quantum entanglement. 
       Recent work has shown that local, random measurements applied to a generic volume-law entanglement generating many-body system are able to force a transition into an area-law phase. This work shows that projective operations can also force a similar classical phase transition; we show that local projections in a chaotic system can freeze information dynamics. In rough analogy with measurement-induced phase transitions, this is characterized by an absence of information spreading instead of entanglement entropy.
       We leverage a damage-spreading model of the classical transition to predict the butterfly velocity of the system both near to and away from the transition point. We map out the full phase diagram and show that the critical point is shifted by local projections, but remains in the directed percolation universality class. We discuss the implication for other classical chaotic many-body systems and the relation to synchronisation transitions. 
	\end{abstract}
	
	\maketitle

\section{Introduction} 
Measurement induced phase transitions (MIPT) are a novel class of dynamical phase transitions which arise when projective measurements which are random in time and space are able to prevent entanglement spreading in a many-body system \cite{PhysRevB.98.205136,PhysRevX.9.031009,PhysRevB.99.224307,PhysRevB.100.134306}.
MIPTs are characterized by a transition between an entangling and disentangling phase, driven only by the addition of random local measurements of the quantum state.
This work examines to what extent such behavior is able to be reproduced in a fully classical system where there is no notion of a measurement or of entanglement entropy.
However, we ask if a classical analog does exist, what is the minimal set of ingredients needed to produce behavior which looks like a MIPT and what is the nature of the transition?

The entanglement entropy of a one-dimensional $N$-site quantum system prepared in a product state is understood to generally grow linearly in time, a phenomenon which has recently attracted study in systems of random unitary circuits \cite{Calabrese_2005,PhysRevX.7.031016,PhysRevB.99.174205}. Such a feature is understood to be due to local entangling from the sequential application of unitary operators, spreading information across the system until at late time the entanglement entropy scales with $N$ --- a volume-law phase in one dimension.
The addition of random non-unitary measurements at a high enough rate counteracts this local spreading of entanglement and has been observed to cause a transition into area-law behavior, where entropy saturates to a constant independent of $N$ and global entanglement growth is stopped.

\begin{figure}[h!]
\centering
\includegraphics[width=0.4\textwidth]{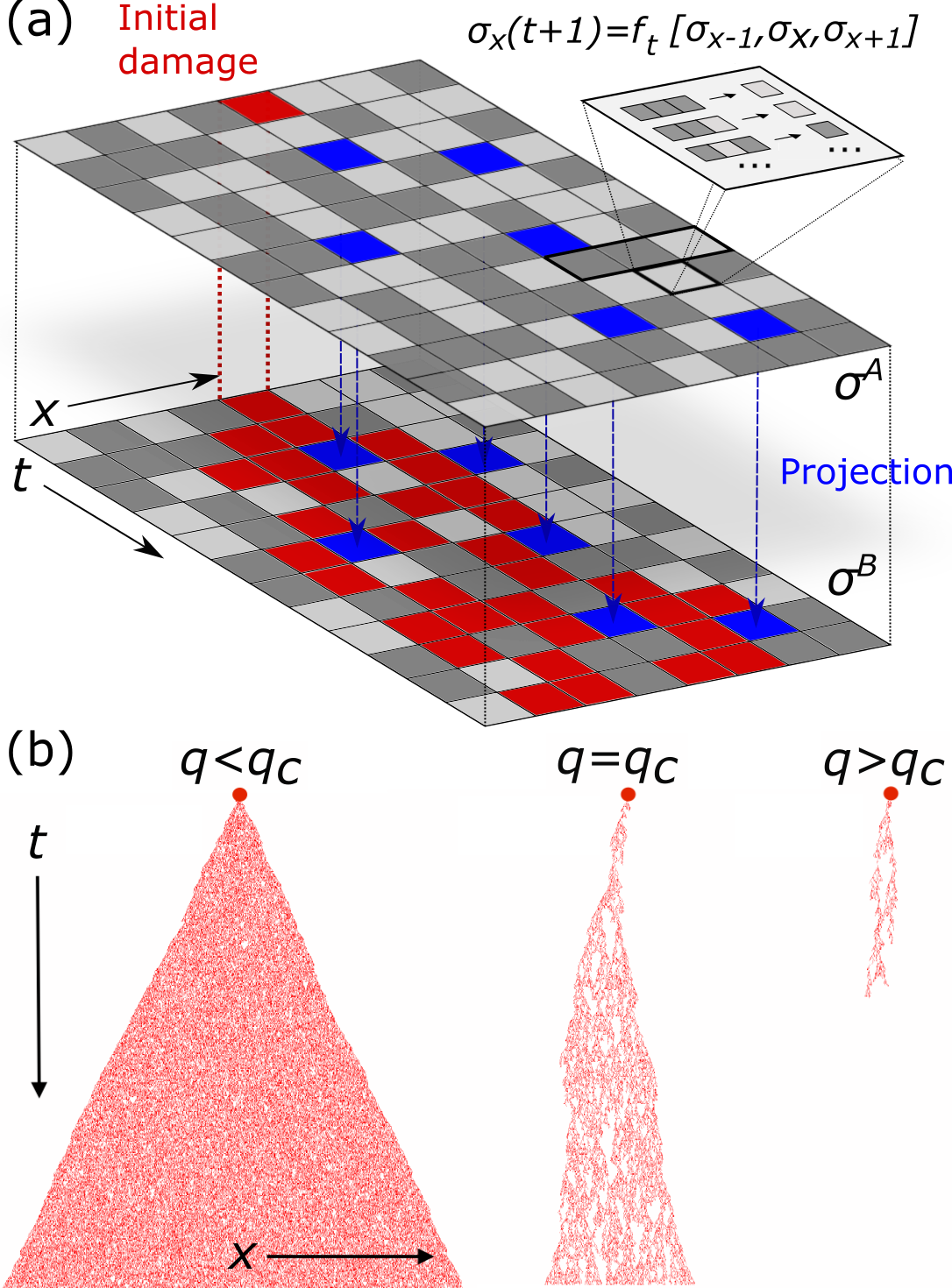}
\caption{(a) Schematic of the Kauffman cellular automaton: a configuration of Boolean elements $\sigma(x,t) = \pm 1$ evolves in time under local rules.
The effects of an initial perturbation is studied with the spaciotemporal decorrelator which measures the difference between two copies differing only by such a perturbation.
Both copies evolve under the same local Kauffman CA rules (as demonstrated in the inset by three such rules at a given position and time), which are randomly generated anew at each timestep.
The decorrelator is represented by colouring sites of copy $B$ which differ from the reference copy $A$; in the chaotic phase of the model, the effect of the perturbation is able to lead to a widening footprint of the decorrelator in space.
The single site with initial damage is additionally marked red in both copies.
Measurements are represented by projecting the reference copy onto the perturbed copy and are shown with blue vertical lines.
(b) Damage-spreading model: the decorrelator defined between two copies of a single realisation of the Kauffman CA, evolving in time. The measurement rate $q$ is tuned to induce a phase transition into the frozen phase.}
\label{fig:schematic}
\end{figure}

How can one identify such a transition in a classical system? In the absence of a classical definition of entanglement entropy, we must instead find another measure which probes how information spreading is suppressed across the transition.
For this purpose we focus on a \emph{damage model} of a classical system, where we quantify how information is scrambled by chaotic dynamics by measuring the local difference between two copies which differ only by a single-site perturbation of the initial state. This local difference is the `decorrelator' due to the perturbation and can be readily applied to many classical chaotic systems to quantify information scrambling via the butterfly effect. 
This chaotic phenomenon still fundamentally represents the spatial propagation of information in the system, which is what we are interested in analogizing from quantum MIPT models.
Here, we focus on a minimal one-dimensional cellular automaton (CA) to uncover classical MIPT-like behavior and discuss its general features.
In the study of cellular automata, the initial perturbation is often called damage, and the spacially-propagating growth of the decorrelator is often referred to as damage spreading --- language which we will adopt here.
Fig.~\ref{fig:schematic}(a) shows the damage spreading due to an initial perturbation in red when this model is in the chaotic phase. The total damage summed over all space, called the Hamming distance, grows linearly in time as the information spreads across the whole system. In previous work \cite{SWLiu2021} we showed that the spatially resolved decorrelator averaged over many-realisations follows a predictable functional form with a velocity-dependent Lyapunov exponent (VDLE) akin to the out-of-time-order correlators (OTOCs) of quantum many-body systems, a newly studied measure of information spreading \cite{2016maldacena,PhysRevB.98.144304,PhysRevLett.122.020603,PhysRevX.8.021014} which, in contrast to the entanglement entropy, has a direct classical analog \cite{PhysRevLett.121.024101,PhysRevLett.121.250602,PhysRevLett.127.124501,10.21468/SciPostPhys.7.2.022,10.21468/SciPostPhys.11.5.087,PhysRevB.103.174302,PhysRevE.102.022130,PhysRevX.8.021013}.

The damage spreading description also provides us with a natural way to incorporate a classical analog of measurements into the system as depicted in Fig.~\ref{fig:schematic}(b): by projecting both copies to be equal at some rate $q$.
In the damage model description a projection sets the decorrelator to be \emph{locally} zero and decreases the distance between the initial and final states in a similar way to the quantum case \cite{PhysRevX.9.031009}, acting against the spread of information, which we expect to drive the system out of a chaotic phase with information propagation and into the frozen phase.
We will henceforth use the term `measurements' to refer to both the quantum and classical local  projections.

Our study is related to recent works employing classical CAs
 for tractable simulations of quantum unitary circuits \cite{PhysRevB.98.060302,PhysRevB.100.214301,PRXQuantum.2.010329,gopalakrishnan2018facilitated,PhysRevLett.127.235301} in which the entanglement entropy can be calculated from copies of classical CAs; we expect that the corresponding OTOCs can serve as an alternative measure with similar behavior because both information-spreading measures are computationally defined as differences between copies evolving under identical rules.
In particular, Iaconis {\it et al.} \cite{PhysRevB.102.224311} have studied MIPTs in such quantum systems, with the mapping to CAs being key in their analysis. The effect of measurements on the growth of entanglement entropy is indeed quantified by setting different copies of the classical CA to be equal.
Given that MIPT are often simulated as classical models, we here address the pertinent question whether one can reproduce and quantify MIPT-like behavior in fully classical many-body systems.

In order to study a typical transition in detail we focus on the maximally stochastic, classical cellular automaton: the annealed Kauffman CA (KCA) \cite{PhysRevLett.84.5660}.
The KCA is defined in terms of local rules mapping a set of binary values onto another, controlled by the parameter $p$. With increasing $p$, the $+1$ state becomes more favoured and (at sufficient connectivity) there is a transition to a chaotic phase, characterized by the growth of the Hamming distance in time. 
It is well known that the transition is in the directed percolation universality (DP) class \cite{doi:10.1142/4016,doi:10.1080/13642818708215325} and a direct mapping with a DP model will be leveraged in this work to describe its critical behavior. 
The classical KCA system has a long history of study across physics, biology and data science and provides a particularly simple platform to study chaotic many-body behavior \cite{Stauffer1987,STAUFFER1988255,Stauffer1994,STAUFFER1989341,Stauffer_1991,KAUFFMAN1969437,KAUFFMAN1969437}.
We find that the addition of projective operations with a rate $q$ are able to qualitatively change the behavior of the decorrelator and completely suppress information spreading, thus realising a classical MIPT.

In this work we focus on the KCA as a model of classical information spreading but we propose that this damage-spreading model can be widely extended to other classical chaotic systems, for example Hamiltonian dynamics, and that local projective measurements will generally lead to the suppression of information spreading. 
This paper is outlined as follows: the KCA system and its simulation are defined in detail, followed by a discussion of its phase diagram in the presence of measurements. Two models are then presented which describe both the behavior away from and at the phase transition, permitting us to relate the system with measurements to the measurement-free system.
Finally, we highlight how it is important to look beyond just entanglement entropy as the measure of MIPTs, and hope to motivate the study of information spreading in other classical and quantum many-body systems using OTOCs.

\section{KCA System and Measures of Information Spreading}  
A local KCA is a system of $N$ Boolean elements $\sigma(x,t)=\pm 1$ which evolve in discrete time steps through local rules that depend upon each site and its $2K$ nearest neighbours in 1D. This process is shown on one of the sheets of Fig.~\ref{fig:schematic}(a, inset). Our KCA system evolves under a set of local rules $\{f_{t}\}$: 
\begin{equation}
    \sigma(x,t+1) = f_{t}\left[\sigma(x-K,t),\dots,\sigma(x,t),\dots,\sigma(x+K,t)\right].
\end{equation}
At any one time step, the cellular automaton evolves under a rules $\{f_{t}\}$ which map $(2K+1)$ inputs to a single output whose value is $+1$ or $-1$, chosen randomly for each input configuration $\{\sigma_{in}\}$ randomly
\begin{equation}
    f_{t}\left[\{\sigma_{in}\}\right] =  \begin{cases} 
      +1 & \textrm{with probability }p \\
      -1 & \textrm{with probability } 1-p.
   \end{cases}
\end{equation}
At any particular time $t$ the same local set of rules is applied to each site across the system --- the same rules always lead to the same output and the evolution is therefore deterministic.
In the annealed KCA system studied here, a new set of rules are randomly chosen at each time step; in numerical simulations observables must be computed by Monte Carlo-sampling many different simulation runs with randomly selected rules for the same $p$.

The probability of any site being $+1$ is constant and equal to $p$ at all time steps, but the transition present in the system is regarding the spreading of information. At low $p$, if a perturbation is introduced at some single site, it is likely that almost all rules do not distinguish between the inputs, and most states are likely to be mapped to $-1$ --- this is the frozen phase of the Kauffman CA. At higher $p$, a single-site perturbation may be expected to lead to an increasing number of perturbed sites after each application of the annealed rules. 
When $K$ is sufficiently large, there exists a critical $p_c$ where under repeated time evolution, the perturbation grows over time and a ballistic propagation of `damage' through the system can occur --- this is the chaotic phase of the KCA.


The tendency of local perturbations to either decay or spread is typically diagnosed with the \emph{global Hamming distance} \cite{Derrida_1986},
\begin{equation}\label{hammdist}
    H(t) = \frac{1}{2N} \left < \sum_{x}|\sigma^{A}(x,t) - \sigma^{B}(x,t)| \right > ,
\end{equation}
between as the fraction of differing sites between two copies of the system $\sigma^{A,B}(x,t)$ which differ by a single inverted site in the initial state at $t=0$.
Numerical studies of the KCA system are based on a Monte Carlo (MC) sampling over the random rules. As illustrated by the two sheets of Fig.~\ref{fig:schematic}, in each simulation instance two copies are instantiated with an identical configuration except one perturbed central site and then propagated under the same rules. The Hamming distance is evaluated via Eq.~\ref{hammdist}, where the average is performed over many MC instances a this given probability $p$.
The distance grows linearly in time in the chaotic phase up to the physical boundary of the system, and decays to zero in the frozen phase, hence mirroring the behavior of the entanglement entropy of quantum many-body phases, see Fig.~\ref{phasediagram}.
In our previous work, we argued that the classical OTOC analog is a spatially resolved {\it local} Hamming distance, or decorrelator, \cite{SWLiu2021}
\begin{equation}\label{decorr}
    D(x,t)  =  \frac{1}{2} \left[ 1 - \langle \sigma^A(x,t)\cdot  \sigma^B(x,t)\rangle \right],
\end{equation}
related to the global distance by $H(t) = N^{-1}\sum_x D(x,t)$. 
The OTOC measures the spacial growth of perturbations in the system, which have been shown to be well described by a velocity-dependent Lyapunov exponent \cite{PhysRevB.98.144304}
\begin{equation}
    D(x=vt, t) \sim \exp(-\mu (v-v_b)^2 t).
\end{equation}
This result holds in the long-time limit, in the chaotic phase sufficiently far from the phase transition (\textit{i.e.} for $p$ sufficiently larger than $p_c$ such that the critical behavior is negligible).
This OTOC is disorder-averaged, but a single-shot instance is used in the mapping to a percolation system and will become important in our study of the critical regime.

The additional ingredient needed to emulate MIPT-like behavior is an implementation of measurements.
We choose here the simplest possible measurement scheme that preserves the occupation of $+1$ states, that is randomly locally fixing sites in $\sigma^B(x,t)$ to be equal to $\sigma^A(x,t)$.
As shown in Fig.~\ref{fig:schematic}, this is implemented numerically by randomly setting the configuration $\sigma_B(x,t)$ with a probability $q$, which serves as our measurement rate, performed prior to considering any of the aforementioned update rules.
This is analogous to making a projective measurement on quantum many-body system, punctuating its unitary evolution. In the Kauffman CA, the parameter $p$ can be viewed as the strength of the information scrambling dynamics that maintain the growth of distance between copies, whereas the parameter $q$ characterizes the collapsing of information that reduces the Hamming distance.

\section{Numerical Results}

\begin{figure*}
\centering
         \includegraphics[width=\linewidth]{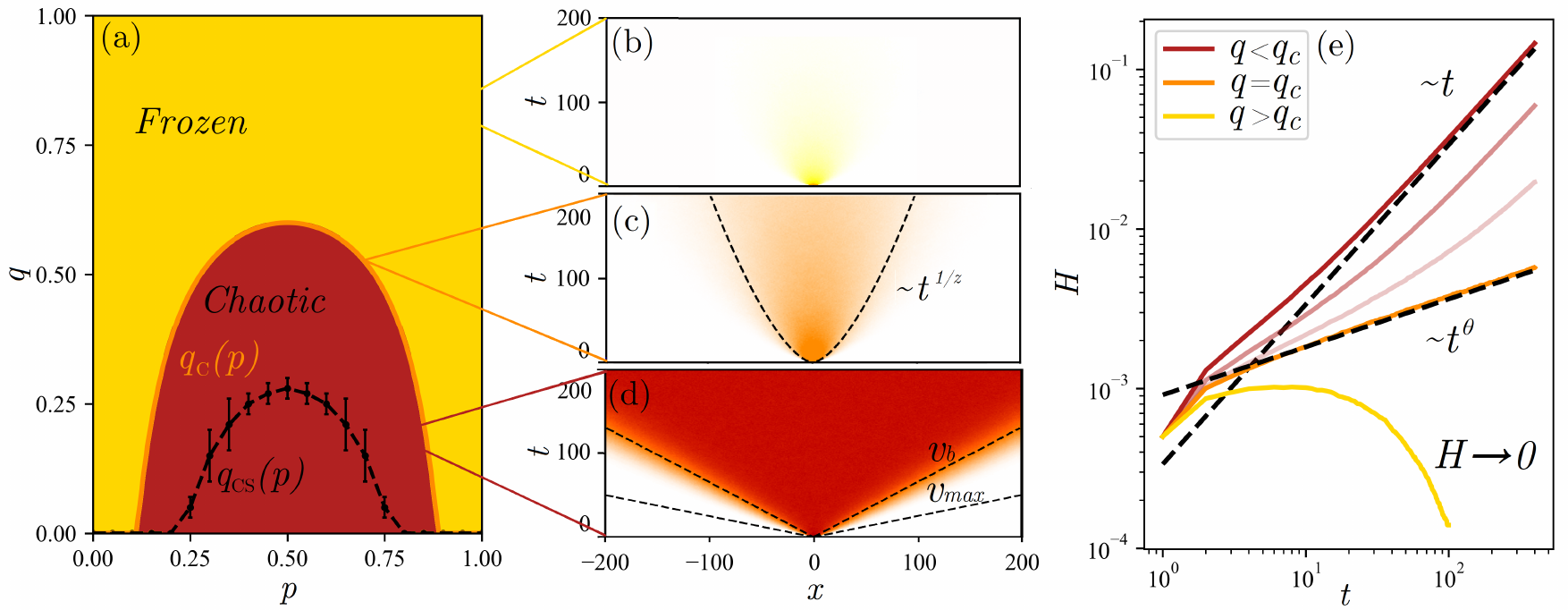}
        \caption{Phase transition of a Kauffman CA with projective measurements diagnosed by the OTOC, with connectivity $K=4$. Left panel: (a) Phase diagram showing the analytic boundary between frozen and chaotic phases. Its functional form is described in the text and confirmed with the data collapse in the critical regime; numerically fixed by precise measurements of the measurement-free system and error bars not shown. The region above the dashed black line $q_{\mathrm{CS}}$ highlights where the observable $v_b$ in the chaotic phase is best described by the critical scaling (CS) DP model \cite{SWLiu2021}; the determination of the crossover point and its corresponding uncertainty is performed using data as plotted in Fig.~\ref{scaling0.4} (see Section~\ref{secdp}). Center panels plot the OTOCs demonstrating at different points of the phase diagram (control parameter $p=0.4$, system size $N=2048$) showing (b) exponential decay of correlations $q>q_c$, (c) critical power-law spreading of correlations $q=q_c$, and (d) lightcone spreading of the OTOC in the chaotic phase ($q<q_c$) at the butterfly velocity $v_b$, which is dependent on the parameters $(p,q)$. This is always less than the maximum velocity at which information could travel $v_{\mathrm{max}}=K$. The right panel (e) plots the Hamming distance $H(t)$ for a range of $q$ with $p=0.4$, $N=2048$.
        It shows linear growth in the chaotic phase, power-law critical growth (with DP exponent $\theta$ \cite{Jensen_1999,PhysRevE.88.042102}) at $q_c$, and exponential decay of perturbations caused by measurements above the critical rate.}
        \label{phasediagram}
\end{figure*}

Fig.~\ref{phasediagram}(e) shows the results of adding local projective measurements which are random in space and time to the Kauffman CA system at a rate $q$: For a fixed $p$, the Hamming distance diagnoses a phase transition at some $q_c$. Above this critical rate the Hamming distance tends to zero in the long time limit and below it grows linearly in time, governed by the butterfly velocity, up to an extensive value.
At the critical point, the Hamming distance grows with the DP critical exponent $\theta \approx 0.3137$ \cite{Jensen_1999,PhysRevE.88.042102}.

The concomitant transition from ballistic propagation to temporal decay of the OTOC is demonstrated in Fig.~\ref{phasediagram}(b--d).
In this work, we show that the suppression of $v_b$ due to projective measurements can be captured by a simple boundary random-walk model, which is also able to predict the functional form of the OTOC deep in the chaotic phase (that is below the dashed lines on Fig~\ref{phasediagram}(a)).

However, as criticality is approached the information spreading is no longer well described by the local random walk of a single boundary site, but an analysis based on the diverging correlation lengths of the model is needed. This occurs because measurements make the perturbed sites critically sparse, and makes it possible that this boundary site (defined as the furthest from the initial perturbation) jumps back inwards by a larger distance than the boundary model predicts.
In this case the mapping of the system to a critical directed percolation model is leveraged to accurately predict the vanishing of $v_b$ as $q\to q_c$ for a range of $p$.
Indeed, given that a mapping to DP is possible for all $p$, we argue that the transition for all $p,q$ is in the same DP universality class.

Taking the critical $p_c$ from the measurement-free system, a mean field treatment can approximate the phase transition line. The expected number of perturbed sites after one application of the rules is $\left<n\right> = p_d(2K+1)$ with $p_d$ the probability of a move, defined below, which should grow in the chaotic phase and reduce in the frozen. 
This constraint can be used to predict the critical line as shown in Fig.~\ref{phasediagram}(a), and is confirmed by the mapping to a critical DP model and numerical data.
This theory is bolstered by its ability to correctly predict the spatial growth of the OTOC and correspondingly the growth of the Hamming distance at criticality with a DP power law (see Fig.~\ref{phasediagram}(c,e)).
A comparison of the critical DP model and the boundary random-walk model is included and we are able to quantify which regions are best described by both models.

\section{Boundary Model of Damage Spreading}  

The adapted KCA system's boundary random-walk model can be used to predict $v_b$ away from criticality \cite{SWLiu2021}.
We define the boundary as the furthest site from the initial damage which differs in the two copies and describe the motion of this site as a random walk.
From understanding the simple rules that govern the boundary movement, the expectation value of the boundary is shown to propagate at a speed $v_b$ and broaden with $\sqrt{t}$ behavior. 

The furthest the boundary could move outwards in one time-step is $K$, and this has probability $2p(1-p)$. We introduce the additional feature that the system is then randomly `measured' with probability $q$, renormalising the $K$-move probability to be $p(K) = p_d = 2p(1-p)(1-q)$.
The probability that the boundary moves fewer steps $x<K$ is therefore $p(x) = p_s^{K-x}p_d$, when $p_s = 1-p_d$.
Therefore the effect of measurements is to reduce the probability of information travelling $p_d$ by an amount proportional to the measurement rate.

\begin{figure}[h!]
\centering
\includegraphics[width=0.5\textwidth]{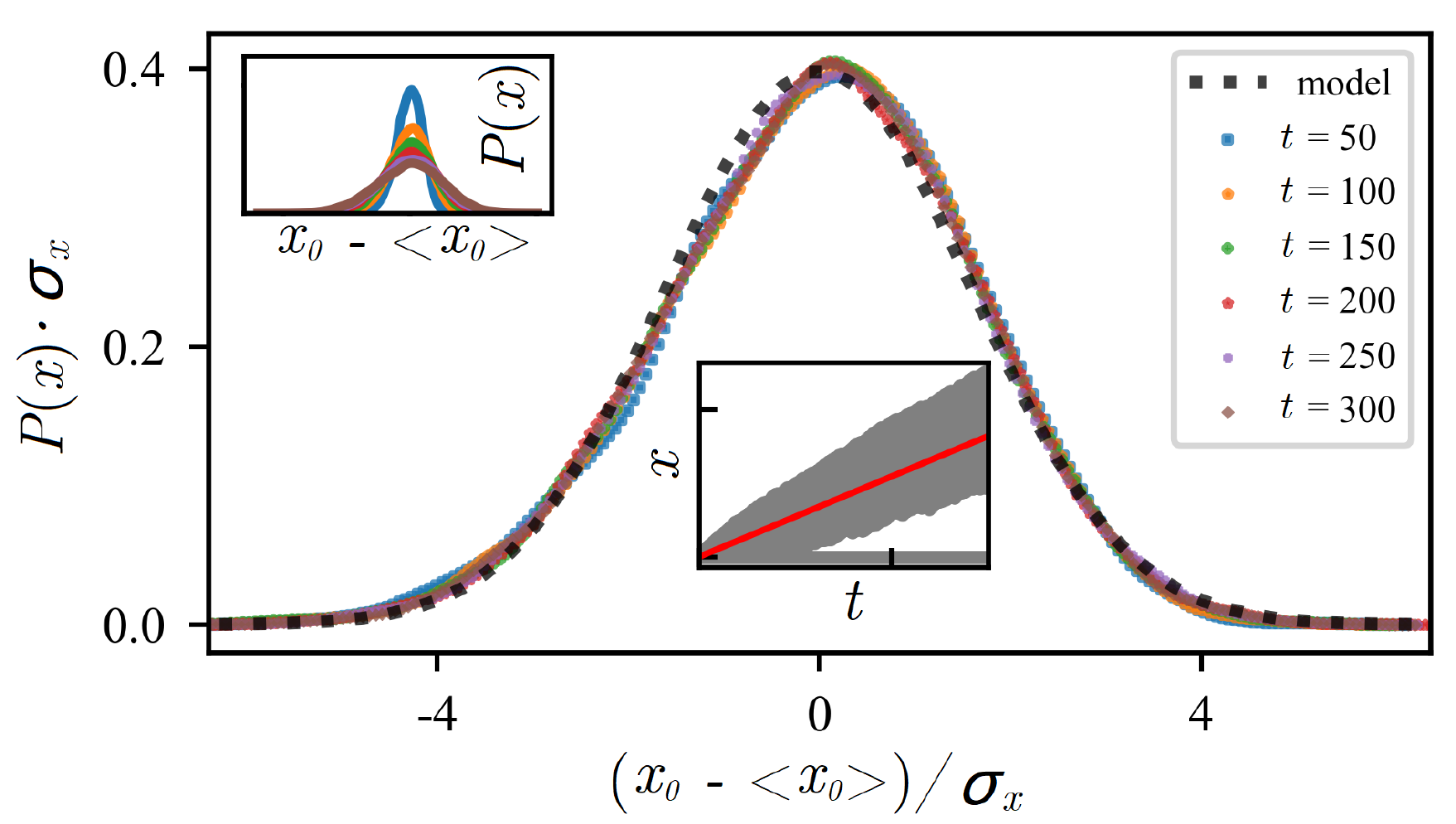}
\caption{Main figure: Scaled boundary distribution for different realizations of the $p=0.4, q=0.4$ model. Top-left inset: unscaled distribution. Bottom: each gray point is the boundary position of one realization, plotted in $x-t$ space. The red line is the averaged boundary position.}
\label{fig:bg_all}
\end{figure}

In Fig.~\ref{fig:bg_all}, we show that even in the presence of the measurements the Gaussian profile of the boundary distribution persists with only a suppressed $v_b$ and an increased variance $\sigma$. 
This analysis is performed by averaging over many starting configurations deep in the chaotic phase, and excluding the few instances where initial measurements remove all damage before it is able to spread (and thus focusing on the long-time behavior).

\begin{figure}[h!]
\centering
\includegraphics[width=0.5\textwidth]{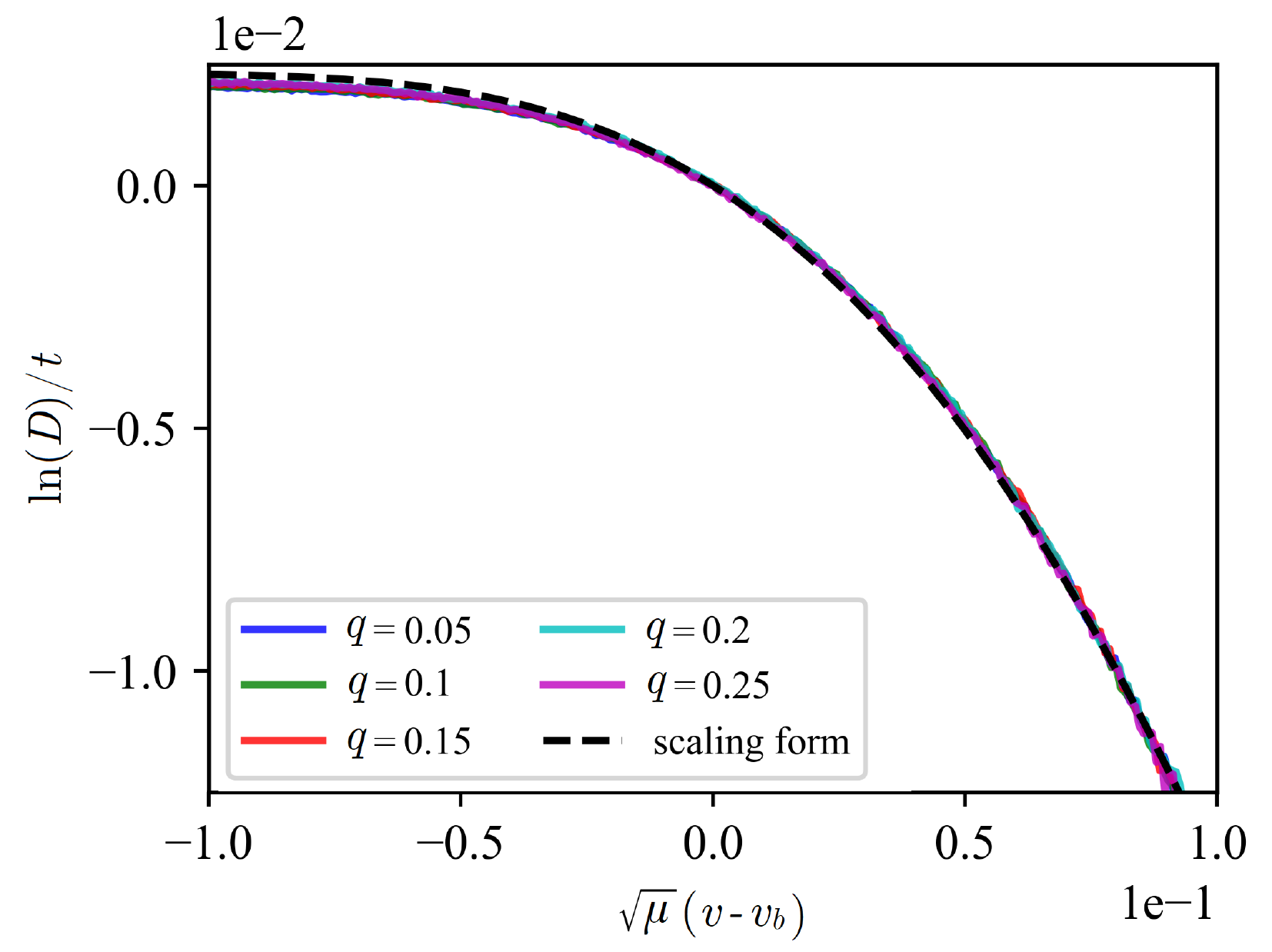}
\caption{Scaling collapse of the numerically calculated decorrelator $D(v,t)$ in the long time limit (colored lines) performed for a range of different meaurement rates $q$ at fixed $p=0.4$. For each specific $q$, data are obtained from the value of $D$ in the long time limit. The horizontal axis is rescaled by $\sqrt{\mu}$ and centered on $v_b$, as calculated from the boundary model accounting for the effect of measurements. The fit line is the analytic scaling form (see main text) \cite{SWLiu2021} which has VDLE exponent $\beta=2$ for $v>v_b$.}
\label{fig:col2}
\end{figure}

The boundary spread can be characterized from moments of the boundary-movement distribution and is well approximated by the central limit theorem in the long time limit \cite{SWLiu2021}.
Furthermore, this model can be applied to predict the leading form of the spatiotemporal decorrelator;
Fig.~\ref{fig:col2} displays a data collapse of $D(v-v_b, t)$ in the large-$t$ limit which demonstrates a good agreement for low $q$.
The collapse is performed by scaling the velocity parameter $\sqrt{\mu} (v-v_b)$, where the parameters are evaluated using the boundary model 
\begin{equation}
    v_b = K - \frac{p_s}{p_d},\quad
    \frac{1}{2\mu^2} = \frac{p_s}{p_d} - \frac{p_s^2}{p_d^2},
\end{equation}
and where $p_{s,d}$ are evaluated using the boundary model including measurements.
Explicitly, the scaling form $\ln{D(v,t)} \sim -\mu(v-v_b)^\beta t$ is reproduced for $v>v_b$ in this system with local projective measurements. The figure shows a good agreement with both the predicted quadratic $\beta=2$ behavior for $v>v_b$, as well as the plateau value inside the lightcone. The full functional form as plotted in Fig.~\ref{fig:col2} takes the form of a Gaussian integral and is numerically evaluated from Ref.~\cite{SWLiu2021}.

\section{Directed Percolation and Critical Damage Spreading}\label{secdp}
Far away from the phase transition, we previously developed a boundary random-walk model of the KCA to predict the characteristic butterfly velocity of the lightcone.
This model assumes that in any system realisation, there is only a narrow physical strip at the boundary of a perturbation in which the decorrelator saturates. By considering the system as a DP model, where differences between the two original copies are mapped to occupied sites, we may explain this intuitively. Away from the transition point, the inside of the percolating system is dense, and therefore the boundary model may assume boundary propagation is dominated by information directly spreading from the sites locally around the boundary.

However, the assumptions of the boundary model breaks down closer to the phase transition. Here the bulk of the system becomes sparse, and it becomes significantly more likely that the previous boundary will not propagate, thus resulting in a new boundary site deep inside the lightcone. As the spatial correlation length $\xi_x$ diverges to be comparable to the system size, this non-local jumping of the boundary dominates and we enter the critical regime.
The previous model takes the perspective that the boundary velocity can be well-described by the local behavior of boundary spreading. This is well founded at late times far from the critical point since all sites in the lightcone are causally connected to the perturbation.
However in the critical regime $p\to p_c$ we expect that finite-size holes form which will causually disconnect a divergent number of sites inside the lightcone from the initial perturbation. In this regime we will use the critical properties of directed percolation to predict the butterfly velocity.


The directed percolation model has two correlation lengths $\xi_{x,t}$ which characterize cluster sizes in temporal and spatial directions \cite{hinrichsen2000non}. At the critical point, these have the following divergent behavior
\begin{equation}
    \xi_\mu \sim |\rho-\rho_c|^{-\nu_\mu},
\end{equation}
where $\rho$ is the percolation probability.
The characteristic velocity of this system is hence given by the dynamical critical exponent $z^{-1}=\nu_x/\nu_t$.

To extend this to the KCA system with measurements, we can draw the connection to percolation through $\rho\to p_d = (1-q)2p (1-p)$, which represents the probability of a site affecting one of its neighbours.
At constant $p$ and under variation of the measurement rate $q$, the critical behavior of the boundary velocity is therefore
\begin{equation}\label{percscalingv}
    v_b \sim |q_c - q|^{z^{-1}}.
\end{equation}
This predicts the that the velocity has critical behavior with the exponent belonging to the directed percolation universality class $z^{-1} \approx 0.6326$ \cite{Jensen_1999,PhysRevE.88.042102}.
The critical measurement rate for any $p$ is hence derived, in terms of the measurement-free critical $p_c$, to be $q_c(p) = 1- p_c(1-p_c)/[p(1-p)]$; thus successfully predicting the form of the phase boundary plotted in the phase diagram Fig.~\ref{phasediagram}(a).

\begin{figure}[h!]
\centering
\includegraphics[width=0.5\textwidth]{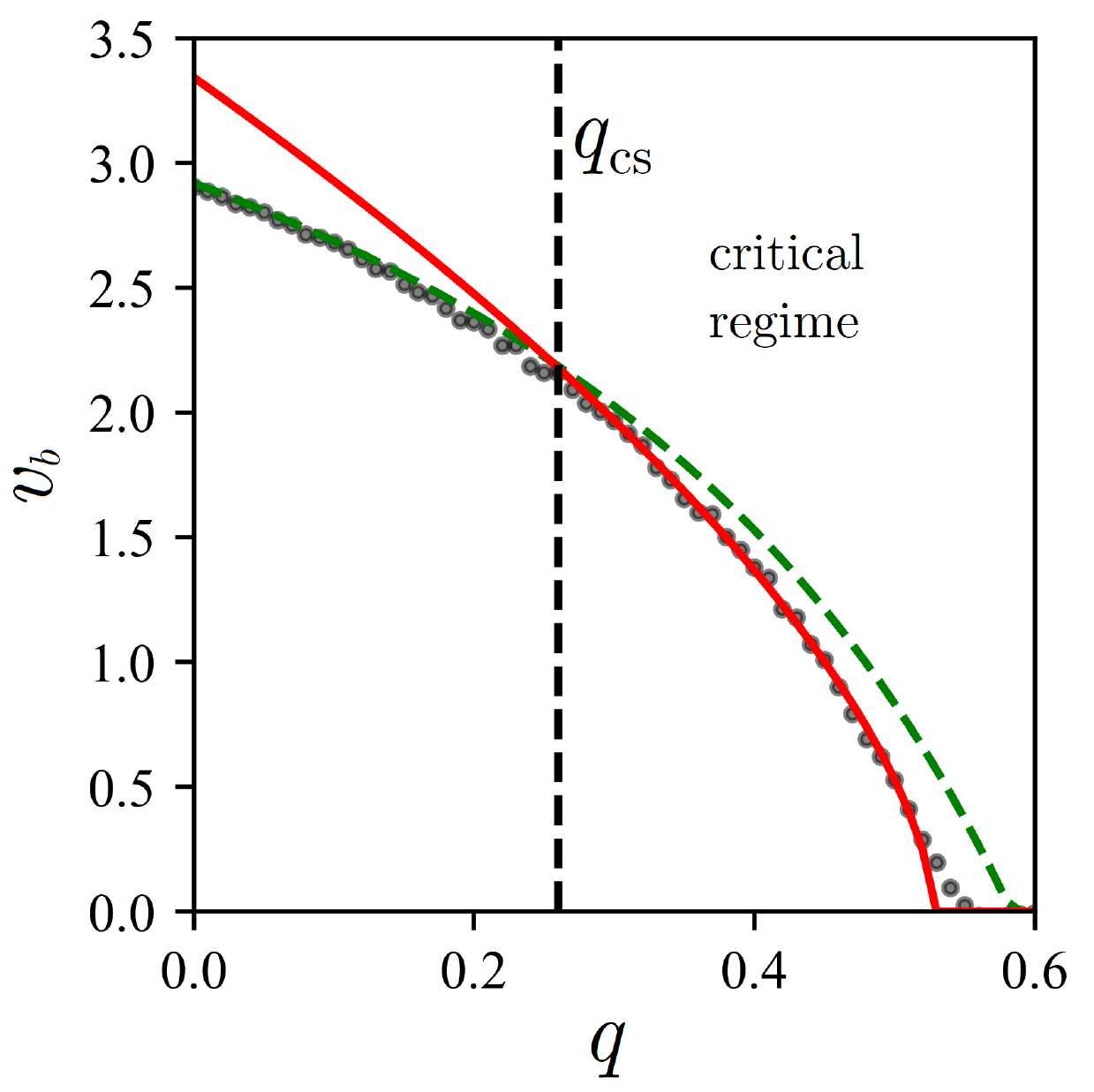}
\caption{Numerical butterfly velocity $v_b$ with varied measurement rate $q$ and fixed $p=0.4$ (grey dots). The data are well described by the boundary model (dashed line) at low $q<q_{\mathrm{CS}}$, and by the critical scaling DP model (solid line) in the range $q_{\mathrm{CS}} \leq q \leq q_c$, with $q_{\mathrm{CS}}$ shown to be approximately $0.3$ in this instance. Data are collected on a $K=4$ system with system size $N=2048$.}
\label{scaling0.4}
\end{figure}

More specifically, taking $v_b=A|\rho-\rho_c|^{z^{-1}}$ below the transition, then the constants $\rho_c$ and $A$ can be fixed by fitting to numerical simulations of the standard $q=0$ system.  
The Fig.~\ref{scaling0.4} shows the predictions of $v_b$ in both the boundary and DP models applied to the case of $p=0.4$ for varying measurement rate $q$. This DP model is able to predict accurately the velocity approaching the critical measurement rate, and even extend the previous work by fitting in the regime of approximately $p<0.3$ where the measurement-free model also becomes critical. 
Together these two models describe the boundary spreading well for the whole chaotic phase; the crossover rate $q_{\mathrm{CS}}$ above which the DP model with critical scaling (CS) predicts $v_b$ better than the boundary model is measured for each individual value of $p$ [see Fig.~\ref{phasediagram}(a)] and is plotted as the black dashed line in Fig.~\ref{phasediagram}.

\begin{figure}[h!]
\centering
\includegraphics[width=0.5\textwidth]{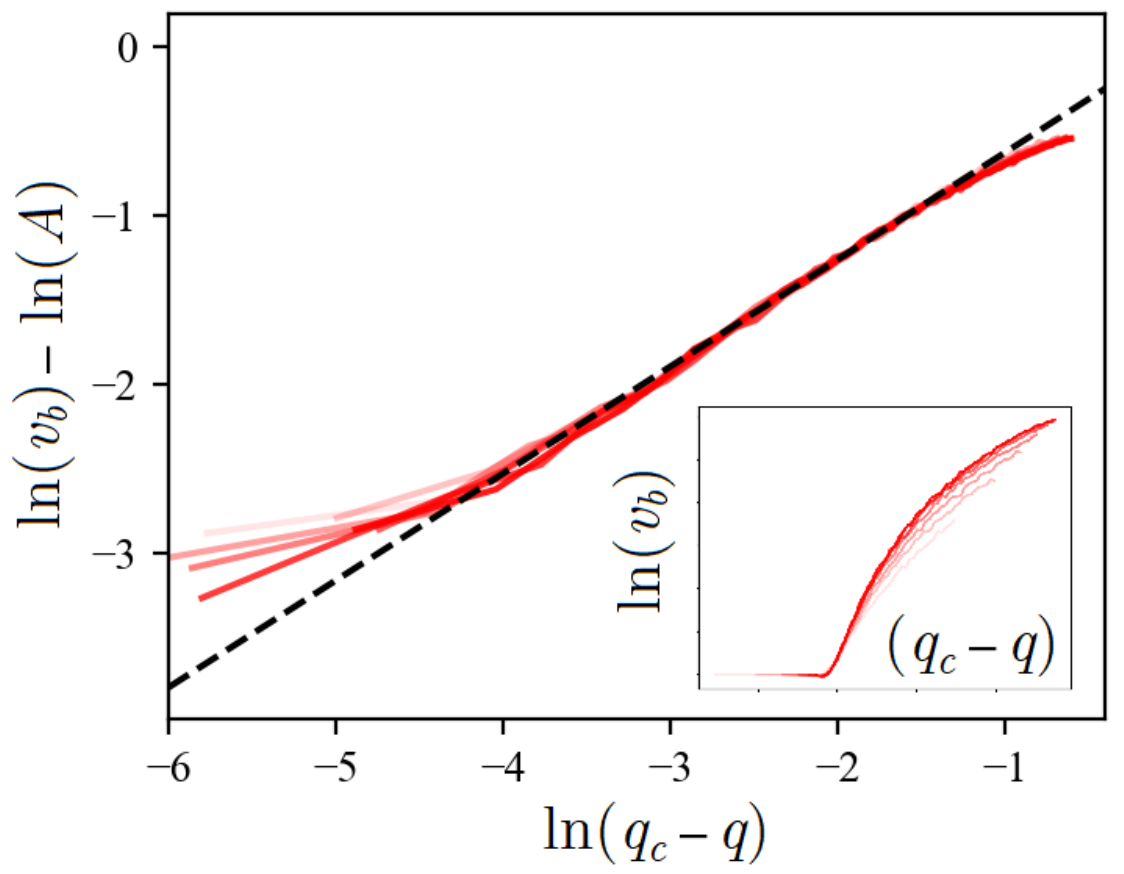}
\caption{Scaling collapse of the butterfly velocity as a function of measurement rate performed for the range $p=0.2$--$0.5$; the dashed line highlights the power-law behavior with critical exponent $z^{-1}$. The logarithm of butterfly velocity is plotted and shifted by the appropriate normalisation $\log(A)$, defined in terms of $q$ in the main text. The horizontal axis is the distance of the measurement rate from the critical rate $q-q_c$. The figure inset shows the same data on a semilog plot and without the data collapse.}
\label{scollapse}
\end{figure}

To confirm numerically our hypothesis that along the phase boundary the model maps onto the same DP system with the same $\rho_c$ and exponents, we perform a data collapse over $p$ using the predicted critical form Eq.~\ref{percscalingv}.
The collapse shown in Fig.~\ref{scollapse} is done by mapping all parameters onto the variables $\rho$, and we find that the data points to a universal transition with common critical behavior; only finite-size effects and the behavior far from $\rho_c$ differ.

\section{Discussion \& Conclusion} 
We have demonstrated that a chaotic classical many-body system can exhibit an information-spreading transition driven only by local projective measurements. Although entanglement entropy is not accessible in classical systems, this poses no hindrance to identifying an information spreading transition.
We have identified a basic damage-spreading model quantified by a decorrelator which is able to measure the evolving spatial impact of a perturbation over time.
This information measure is the local Hamming distance between two copies of our CA system which initially differ at a single site. When disorder averaged it can be thought of as the classical analog of an OTOC. As one may expect, we find that the velocity-dependent Lyapunov exponent behavior of the OTOC is unaffected by local measurements, but the velocity of information spreading is parametrically suppressed.


Our work raises the question of what behavior would an OTOC demonstrate when applied to a quantum MIPT? It is unclear how the presence of random, local, non-unitary measurements across the system would affect the ability of two operators at different points in space-time to commute. Indeed this question seems to have no consistent understanding in the MIPT literature.
Nevertheless, the ability to describe quantum information-spreading transitions with OTOCs would provide another important tool to understand to what extent quantum behavior can be analogized classically.

From our results on MIPTs in a minimal CA we expect that such transitions can appear generically in other paradigmatic classical models. 
For example, it is understood that a chaotic classical Heisenberg chain shows information spreading which is also described by an OTOC with VDLE behavior \cite{PhysRevLett.121.024101}.
The OTOC is similarly defined as a disorder-averaged damage model, thus, projecting  two copies of the spin model onto a common basis state at a sufficient rate $q_c$ is also  expected to prevent any information spreading in the system.
A perhaps even simpler realization could be  coupled classical oscillators, with the benefit that there is a one-dimensional state space versus the two-dimensional one of the Heisenberg system \cite{classicaltimecryst}. Alternatively, it would be worthwhile to study the effect of kinetic constraints, which lead to distinct classical OTOC behavior~\cite{deger2022arresting}, on classical MIPTs.

It turns out that the tendency of many-body systems to converge due to local projections has been observed before in the context of synchronisation transitions \cite{austenLamacraft}, whereby two random chaotic systems coupled sufficiently strongly `synchronise' and become equal over time \cite{PhysRevE.63.036226,PhysRevLett.88.254101}.
For example, two CA systems $x,y$ evolving under the same local rules will synchronise if one of the systems $y$ is randomly updated from the configuration $x$ at a sufficient rate $q$ \cite{PhysRevE.59.R1307}. Above this rate, and for any initial conditions of the two systems, the two copies will converge at long times. Indeed this has also been studied for KCA systems, using a symmetric coupling scheme in Refs.~\cite{PhysRevE.63.036204,doi:10.1142/S0218127403007114} which also predicts a $p$-dependent critical coupling $q_c$.
Out work complements this literature by investigating the synchronisation of copies which differ only by a single initially damaged site, allowing us to study the spacio-temporal profile of the competition between synchronisation and chaotic dynamics.
These results are naturally provided by the OTOC analog in our system and, as motivated above, we believe this will also allow the local study of synchronisation transitions in other many-body systems.

%
%

The classical OTOC analogue is not the only measure which could be used to identify a classical MIPT but it allowed for a particular simple local projective measurement protocol and analytical understanding. For quantum many body systems the OTOC misses information contained in the quantum entanglement measure. Similarly, a classical measure akin to the quantum entanglement entropy would be helpful for generalising the idea of MIPT to classical systems~\cite{Pizzi2022}. Nevertheless,  the basic idea that local projective operations can counteract the scrambling of information from the intrinsic chaotic dynamics should carry over.

It is an important question to understand whether the critical behavior of different MIPTs also lie in the usual class of DP chaos-spreading transitions.
In our example the MIPT transition is of the same class as the active-frozen transition of the measurement-free system. However it does not exclude the interesting possibility that the nature of the transition changes, or that an intermediate phase (for example with sub-ballistic information spreading) could also exist in classical systems. In addition, different measurement protocols in models with higher dimensional state space could lead to distinct MIPTs.




\section{Acknowledgements}
We would like to thank Subhro Bhattacharjee, Thomas Bilitewski, Johannes Feldmeier, Andrea Pizzi, Jonathan Ruhman, Brian Skinner, and Hongzheng Zhao for their contribution to this work through discussions.
We also thank Austen Lamacraft for bringing to our attention the connection to synchronisation transitions.
This work was in part supported by the Deutsche Forschungsgemeinschaft under grants SFB 1143 (project-id 247310070) and the cluster of excellence ct.qmat (EXC 2147, project-id 390858490) as well as by DARPA via the DRINQS program.

\bibliographystyle{apsrev4-2}
\bibliography{refs}

\end{document}